\documentclass[12pt, a4paper]{article}
\usepackage[utf8]{inputenc}
\usepackage{dcolumn,lscape}
\usepackage{amsmath,longtable,multicol,dcolumn,tabularx,graphicx,amssymb}
\usepackage{exscale,amsthm,multirow,rotating,subcaption}
\usepackage{natbib}
\usepackage[table]{xcolor}
\usepackage{bbm}
\usepackage{hyperref}
\usepackage{tikz,dsfont,float}
\definecolor{col1}{HTML}{EBEBDE} 
\definecolor{col2}{HTML}{777764} 
\definecolor{col3}{HTML}{4F4747} 

\usetikzlibrary{patterns}

\usepackage{algorithm}
\usepackage{algpseudocode}
\begin{document}

\def\spacingset#1{\renewcommand{\baselinestretch}%
{#1}\small\normalsize} \spacingset{1}


\title{Adaptive LASSO estimation for functional hidden dynamic geostatistical models}
\author{Paolo Maranzano\footnote{paolo.maranzano@unimib.it},  Philipp Otto, Alessandro Fassò\\
{\small University of Milano-Bicocca, Leibniz University Hannover, University of Bergamo}}

\maketitle

\begin{abstract}
We propose a novel model selection algorithm based on a penalized maximum likelihood estimator (PMLE) for functional hidden dynamic geostatistical models (f-HDGM). These models employ a classic mixed-effect regression structure with embedded spatiotemporal dynamics to model georeferenced data observed in a functional domain. Thus, the parameters of interest are functions across this domain. The algorithm simultaneously selects the relevant spline basis functions and regressors that are used to model the fixed-effects relationship between the response variable and the covariates. In this way, it automatically shrinks to zero irrelevant parts of the functional coefficients or the entire effect of irrelevant regressors. The algorithm is based on iterative optimisation and uses an adaptive least absolute shrinkage and selector operator (LASSO) penalty function, wherein the weights are obtained by the unpenalised f-HDGM maximum-likelihood estimators. The computational burden of maximisation is drastically reduced by a local quadratic approximation of the likelihood. Through a Monte Carlo simulation study, we analysed the performance of the algorithm under different scenarios, including strong correlations among the regressors. We showed that the penalised estimator outperformed the unpenalised estimator in all the cases we considered. We applied the algorithm to a real case study in which the recording of the hourly nitrogen dioxide concentrations in the Lombardy region in Italy was modelled as a functional process with several weather and land cover covariates.
\end{abstract}

\noindent%

\spacingset{1.45} 

\newpage


\section{Introduction}

With the growing availability of geo-referenced data in high spatial and temporal resolutions, geostatistical applications increasingly require efficient algorithms to select relevant regressors among a large set of candidates.
In this context, statistical methods for such high-resolution spatial data often suffer from the so-called \textit{Big-}N\textit{-problem}, in which the time complexity of estimation algorithms grows polynomially with an order greater than 2 when the number of locations is increasing and traditional methods are often not computationally feasible \citep[cf.][]{katzfuss2017multi,katzfuss2011spatio}. 
To reduce the complexity of such models, various approaches have been used, some of which are based on
inducing sparsity in the spatial covariance matrix \citep{furrer2006covariance,kaufman2008covariance,stein2013statistical,furrer2016asymptotic}.
Some other approaches are related to the precision matrix, either using a graphical least absolute shrinkage and selection operator (LASSO) approach \citep{krock2021nonstationary,krock2021modeling}, or a sparse Cholesky factors approach based on the Vecchia approximations \citep{stein2004approximating,kang2021correlation,schaefer2021sparse} and on multi-resolution approximations of Gaussian processes \citep{Katzfuss2017,JurekKatzfuss2021}. In particular, Vecchia approximation can be efficently used to peform high-dimensional spatiotemporal filtering \citep{JurekKatzfuss2022_StatComp} and spatiotemporal smoothing \citep{JurekKatzfuss2022_Envir}.
Low-rank covariance matrices have been also considered, including fixed-rank kriging and penalised methods \citep{banerjee2008gaussian,cressie2008fixed,chang2010semiparametric,hsu2012group,cressie2010fixed}. Eventually, combined approaches, like the so-called \textit{full-scale approximation of the covariance matrix} have been proposed \citep{sang2012full}.



In addition, the spatiotemporal data may be defined in a functional domain because of their characteristics, e.g., vertical atmospheric profiles in climate studies, \citep{fasso2018statistical}; off-shore coastal profile measurements for beach monitoring, \citep{Otto21_CEng}. A functional data approach may also be used to reduce the dimensionality of high-frequency temporal observations. For example, \cite{ignaccolo2008analysis} considered the time series of air quality measurements at many stations as functional observations. Also, to understand the bike-sharing system, \cite{Piter20_arxiv_Helsinki} considered daily 5-min usage profiles of a bike-sharing system as daily functional observations. Due to the spatial nature of the underlying process, further applications can be found in environmetrics \citep[e.g.,][]{franco2017bootstrap,ignaccolo2013functional,giraldo2011ordinary}, medicine \citep[e.g.,][]{aristizabal2019analysis}, econometrics \citep[e.g.,][]{pineda2019functional}.

For joint estimation and model selection, we will consider penalised estimation procedures. In general, the use of penalised regression approaches for the selection of relevant variables has the advantage of greater stability of the solutions than in the classic backward and forward selection methods \citep{Breiman1996,BondellEtAl2010}. In this regard, we refer to the review paper of \cite{MullerWelsh2010} on model selection curves, in which (1) different loss functions and penalty terms are extensively discussed for the selection of relevant covariates and (2) the loss functions are studied as functions of the penalty term, which allows for the exploration of their stability, instead of evaluating the function at single values of the penalty multiplier.

Furthermore, methods of selecting covariates have been developed based on penalised methods in spatial and spatiotemporal settings. For instance, \cite{WangZhu2009} suggested a penalised least squares approach for spatial regression models; \cite{CaiMaiti2020}, for spatial autoregressive models; and \cite{GonettaEtAl2022}, for conditional autoregressive models. For additive spatial models including potential nonlinear effects, \cite{NandyEtAl2017} developed a weighted penalised least squares estimator. Alternatively, penalised maximum likelihood estimators (PMLE) are considered. For instance, \cite{ZhuHuangReyes2010} suggested PMLE for linear models with spatially correlated errors. \cite{ChuZhuWang2011} and \cite{chu2011penalized} additionally reduced the model's complexity by combining covariance tapering and a PMLE for spatial and spatiotemporal settings, respectively.

Following the least absolute shrinkage and selector operator (LASSO) methodology \citep{tibshirani1996regression}, \cite{ReyesEtAl2012} proposed a spatiotemporal adaptive LASSO algorithm for linear regression models with spatiotemporal neighbourhood structures. The estimation strategy involved both the penalised least squares and PMLE approaches. Other examples of penalised regression for spatiotemporal data are in \cite{AlSulamiEtAl2019}, in which an adaptive LASSO method was developed to simultaneously identify and estimate spatiotemporal lag interactions in the context of a data-driven semiparametric nonlinear model. Furthermore, \cite{SafikhaniEtAl2020} considered LASSO methods for generalised spatiotemporal autoregressive models. The estimators are obtained by a modified version of the penalised least squares that accommodates hierarchical group LASSO-type penalties.
In addition to these contributions, there are several application-oriented papers that combine classic LASSO approaches and geostatistical models in multi-step procedures \citep[e.g.,][]{fasso2021spatiotemporal,ye2011sparse,pejovic2018sparse}.

Penalised methods are also commonly applied in the context of functional data analysis, such as penalised splines \citep[see][]{RamsaySilverman2002,claeskens2009asymptotic}. These methods usually regularise the smoothness of the estimated functions by penalising the integrated second derivatives. In this way, many basis functions can be used, thus avoiding the typical overfit resulting from unpenalised estimation methods. In this paper, we will not consider these smoothness penalties, but we will focus on model selection for spatiotemporal models. In general, spline basis functions are widely used tools in geostatistics for the spatiotemporal interpolation of environmental phenomena  \citep[see][for group-LASSO approaches in this context]{HofierkaEtAl2002,XiaoEtAl2016,ChangHsuHuang2010,HsuChangHuang2012}. Also, spatial and spatiotemporal model selection has been addressed in a Bayesian framework  
\citep[see e.g.,][]{katzfuss2012bayesian,CarrollEtAl2016a,CarrollEtAl2016b,LawsonEtAl2017,CarrollEtAl2018} but it will not be the focus of this paper.


In this paper, we efficiently select the relevant regressors of the functional spatiotemporal model known as the \textit{functional hidden dynamics geostatistical model} (f-HDGM). This model decomposes the random process into a fixed-effects part (that includes external regressors) and a random-effects part. The latter component separates spatial and temporal dependence into two different terms, namely a stationary geostatistical process and a Markovian temporal process. That is, the spatiotemporal dependence must be separable (see \cite{huang2007model} for a comparison of different spatiotemporal models). In a multivariate framework, the model and, in particular, the maximum-likelihood estimation procedure were introduced by \cite{calculli2015maximum}. The procedure is based on maximising the expected likelihood function with respect to the parameters using an expectation-maximization (EM) algorithm. This approach is based on the so-called \textit{state-space representation} and the related Kalman Filter algorithm. For a recent review of Kalman filtering for spatiotemporal models, see \cite{FerreiraMateuPorcu2022}, \cite{JurekKatzfuss2021}, and \cite{JurekKatzfuss2022_Envir}. Also, extensions for non linear and non Gaussian spatiotemporal data, such as the Ensemble Kalman Filter, have been considered by \cite{KatzfussStroudWikle2016,KatzfussStroudWikle2020}. The multivariate HDGM approach was then generalised by \cite{finazzi2020dstem2} to functional spatiotemporal models, which are implemented in the MATLAB software package D-STEM.

We propose a LASSO procedure for the selection of the functional coefficients of f-HDGM. 
More precisely, the suggested penalised maximum-likelihood approach with an adaptive LASSO penalty simultaneously estimates the functional coefficients and selects relevant regressors. 
In the functional setting, these variables are the regressive coefficients of the spline bases. Thus, in addition to fully including or excluding regressors, we also find the relevant parts (for the B-spline bases) or frequencies (for the Fourier bases) when only some of the functional coefficients are set to zero. For instance, \cite{Otto21_CEng} showed that major storm floods have an effect only on specific parts of the coastal profiles, that is, those affected by high waves during a flood.

It is important to note that geostatistical applications are prone to cross-correlated regressors due to their spatial nature. As \cite{zhao2006model} pointed out, the classic LASSO approach would generally not be selection-consistent in this case. Also, \cite{SimonTibshirani2012} showed that when the covariates are correlated, the group-LASSO estimator (cf. \cite{YuanLin2006}), which assumes orthonormal data within each group, performs poorly in selecting the relevant features. This motivated the choice of an adaptive LASSO penalty, which led to selection-consistent estimators in the case of cross-correlated regressors \citep[see][]{zou2006adaptive} and \cite{zou2008one}.

The aforementioned method was especially promoted by the Agrimonia project\linebreak (\href{https://agrimonia.net/}{https://agrimonia.net/}), which aims at building spatiotemporal models for the relation between livestock and air pollution. In such a contexts, the number of regressors, which includes meteorology, land use, socio-economic variables, and their interactions, may become quite large.




The remainder of the paper is structured as follows. In Section \ref{sec:fHDGM}, we briefly introduce the considered functional geostatistical model. In Section \ref{sec:lasso}, we present the new penalised maximum likelihood approach using an adaptive LASSO penalty. In Section \ref{sec:MC}, we present the results of an extensive Monte Carlo simulation of three simulation settings. In Section \ref{sec:AQLombardy}, we illustrate how our proposed estimation algorithm can be used to model daily air quality profiles in Lombardy, a region in Northern Italy, during the coronavirus disease 2019 (COVID-19) pandemic. Finally, Section \ref{sec:conclusion} concludes this paper and identifies potential topics for future research.

\section{The functional model} \label{sec:fHDGM}
The f-HDGM and the corresponding software D-STEM \citep{fasso2018statistical} are designed to handle functional data $\{Y_{s,t}(h): s \in D, t = 1, \ldots, T\}$ defined on the interval $H = [h_1,h_2]$. That is, $Y_{s,t}(h) : H \rightarrow \mathds{R}$ can be observed at any $h \in H$ for any given location $s$ in the spatial domain $D$ and for any given discrete time $t$. Although the spatial domain $D$ is continuous, we will observe such data on $n$ spatial points in an irregular grid $S=\{s_1,...,s_n\}$. Similarly, we will observe the data for each function at a discrete set of points $h_1,...,h_q$, where both $h_i$ and $q$ may depend on $s_i$ and $t$. These observations are denoted by vectors $\{\boldsymbol{y}_{s,t}=(Y_{s,t}(h_1),...,Y_{s,t}(h_q)\}$.
Our data set will composed of $N = nT$ functional data.

To account for the spatial and temporal dependence, we model the process using a hidden dynamics geostatistical process that separates all regressive effects from the spatiotemporal interrelations. More precisely, f-HDGM is defined by
\begin{equation} \label{eq:fHDGM}
    Y_{s,t}(h) = \mu_{s,t}(h) + \omega_{s,t}(h) + \varepsilon_{s,t}(h) \quad ,
\end{equation}
where the fixed-effects component $\mu_{s,t}(h)$, the random-effects component $\omega_{s,t}(h)$ and the modelling errors variance $\sigma^2(h) = VAR[\varepsilon_{s,t}(h)]$ are modelled using splines.
Let $B_{k,a}(h)$ be the $k$-th of the $K_a$ basis functions of component $a = \mu, \omega, \sigma$. 

In Equation \ref{eq:fHDGM}, the mean, or the fixed-effects component, is a linear regression model in the functional domain. That is,
\begin{equation}
    \mu_{s,t}(h) = \sum_{j = 1}^{p}\sum_{k = 1}^{K_\mu} X_{s,t,j}(h) B_{k,\mu}(h) \beta_{jk} \quad ,
\end{equation}
where $X_{s,t,j}(h)$ denotes the functional observations of the $j$-th regressor. For the generic $j$-th regressor, by multiplying of the spline basis matrix by the coefficients $\beta_j = (\beta_{j1},\ldots,\beta_{jK_\mu})'$, we obtain the functional coefficients shown in Figure \ref{fig:TheorCubicSpline}. In Section \ref{sec:lasso}, we propose an adaptive LASSO procedure to penalize these regression coefficients. In a nutshell, whether all entries of the vector $\beta_j$ are shrunk to zero or not, we can select the relevant regressors. That is if $\beta_j$ contains only zeros, then, the $j$-th regressor is removed from the model. Moreover, if $\beta_j$ is only partly shrunk to zero, we can select the relevant parts and knots of the $j$-th regressor in the functional domain. 

In Equation \ref{eq:fHDGM}, the spatiotemporal dependence is modelled by the functional random effects $\omega_{s,t}(h)$, given by
\begin{equation} \label{eq:RE}
    \omega_{s,t}(h) = \sum_{k = 1}^{K_\omega} B_{k,\omega}(h) z_{s,t,k} \, .
\end{equation}
In Equation \ref{eq:RE}, the latent component $z_{s,t} = (z_{s,t,1},\ldots,z_{s,t,K_{\omega}})$ follows a temporal Markovian process, as follows:
\begin{equation}
     z_{s,t} =  G z_{s,t-1} + \eta_{s,t} \quad ,
\end{equation}
with a spatially correlated $K_\omega$-dimensional Gaussian Process $\eta_{s,t}$. In this way, temporal dependence is separated from spatial dependence, which is included in $\eta_{s,t}$, while temporal effects are covered by the transition matrix $G$. The spatial Gaussian process is assumed to have a zero mean and diagonal covariance matrix:
\begin{equation}
     diag(v_1\rho({s}-{s}^\prime ; \theta_1), \ldots , v_{K_\omega}\rho({s}-{s}^\prime ; \theta_{K_\omega}))
\end{equation}
where $\rho$ is a valid spatial correlation function for locations $s$ and $s'$ based on an exponential model with parameters $\{\theta_1,\ldots,\theta_{K_\omega}\}$ and scaling factors $\{v_1,\ldots,v_{K_\omega}\}$, that is, the variances. Both $G$ and $V(\eta_{s,t})$ are assumed to be diagonal, to reduce complexity (i.e., by removing the temporal dependence between different parts along the functional domain).

Eventually, the model errors are assumed to be independent and identically normally distributed across space and time, but the error variance may vary across the functional domain as follows:
\begin{equation}
    \sigma^2(h) = \sum_{k = 1}^{K_\sigma} B_{k,\sigma}(h) \sigma^2_{k} \, .
\end{equation}

Let $\beta = (\beta_1, ..., \beta_p)'$ be the stacked vector of the spline coefficient vectors of the fixed-effects model and $\vartheta = \{G, V, \theta, \nu, \sigma^2\}$ be the set of all coefficients of the random effects and the error term. Moreover, let $H$ denote the Hessian matrix of the model's log-likelihood. The full set of parameters $(\beta, \vartheta, H)'$ is estimated by maximising the expected log-likelihood of the model using an EM algorithm. Let $(\beta_0, \vartheta_0)'$ denote the maximum likelihood estimates of $(\beta, \vartheta)'$, and let $H_0 = H(\beta_0, \vartheta_0)$ denote the Hessian matrix computed at the ML solutions $(\beta_0, \vartheta_0)'$. Notice that the estimates $\beta_0$ is a consistent estimator of $\beta$ \citep[see][]{calculli2015maximum}. The EM algorithm used for computation is implemented in the D-STEM software \citep{Finazzi14,finazzi2020dstem2} within the Matlab environment.

Estimating the parameters of f-HDGM can be computationally demanding. Following \cite{finazzi2020dstem2}, we reduced the computational time with the following two approximations. In the first approximation the variance-covariance matrix of the parameters is computed using an approximated approach. This task is performed by fixing a threshold for the overall improvement in the variance-covariance matrix computation \citep[see Section 2.5 of][]{finazzi2020dstem2}. The second approximation concerns a spatial partitioning approach. According to \cite{stein2013statistical}, we divided the complete dataset into $k$ groups (based on the geodesic distance) of size $r$, and we assumed that the data in the different groups were not correlated. This implies a factorised likelihood function and the possibility of computing the E-step in parallel. As a result, the computational complexity was reduced from $O(Tn^3b^3)$ to $O(Tkr^3b^3)$ \citep[see Section 2.4 of][]{finazzi2020dstem2}. Moreover, if the computing infrastructure can handle $k$ parallel processes, the computing time may be reduced to $O(Tr^3b^3)$.

\section{Spatiotemporal adaptive LASSO estimation for functional coefficients}\label{sec:lasso}
In this section, we suggest an adaptive LASSO approach to select (1) the relevant regressors, (2) the relevant sections of the functional coefficients and (3) the relevant knots of the fixed-effects functional model $\mu_{s,t}(h)$. The emphasis is on modelling the relationship between the covariates and the response variable. Therefore, the parameters of the random-effects components are kept unpenalised. Moreover, for regularised regression approaches, the covariance matrix of the model errors is usually not penalised \citep[e.g.,][]{fan2001variable,tibshirani1996regression}.
Spatiotemporal parameters could also be included in the penalised procedure \citep[e.g., see][for random-effect shrinking in linear mixed models]{BondellEtAl2010}. However, in this case, the shrinkage target should be adjusted to the specific empirical case. Indeed, while the temporal dependence parameter matrix $G$ could be shrunk to zero, i.e., in the case of temporal independence, a zero shrinkage target is not meaningful for the variance parameters and the range parameter of the spatial dependence $\theta$. Regarding the intercept or constant term of the model, we follow the strategy originally proposed by \cite{tibshirani1996regression}, of standardising all the covariates and centering the dependent variable before applying the penalised regression. Such pre-processing step allows for the omission of the intercept term in the adaptive LASSO optimisation  \citep[see Section 2.2 of][]{HastieSLS2015}. Note that, without standardisation, the penalised solutions would depend on the original units used to measure the predictors. Thus, standardisation ensures that the penalty is applied equally to all predictors in terms of the unit variance of all the predictors. Moreover, the estimates can easily be back-transformed with the sample mean and covariance matrix that were used for the standardisation \citep{Lee2015}. As we are dealing with spatiotemporal data, we standardised both the response variable and the covariates with respect to their overall mean and overall standard deviation. That is, we standardised the observations using the 24-hour sample mean and sample standard deviation.


We followed a penalised maximum-likelihood estimation strategy for the fixed-effect coefficients using the following equation:
\begin{equation}\label{eq:adaptiveLASSO_0}
 \beta^{(PMLE)}(\lambda) = \arg \max_{\beta} \mathcal{L}(\beta, \theta) - \lambda f(\beta) 
\end{equation}
with the logarithmic likelihood function $\mathcal{L}$ and a penalty function $f$.
To reduce the computational burden, we locally approximated the full model log-likelihood in (\ref{eq:adaptiveLASSO_0}) around the unpenalised ML estimates using a local quadratic approximation \citep{JennrichEtAl1976,Longford1987}, as follows:
\begin{equation}\label{eq:LQA}
\mathcal{L}(\beta) \cong \mathcal{L}(\beta_0) + \nabla \mathcal{L}(\beta_0)'(\beta - \beta_0) + \frac{1}{2} (\beta - \beta_0)' H_0 (\beta - \beta_0) \, ,
\end{equation}
where $\nabla \mathcal{L}(\beta_0)$ is the score function evaluated at the ML solution and $H_0 = \nabla^2 \mathcal{L}(\beta_0)$. Similar computational solutions involving local approximation of the likelihood have been proposed by \cite{zou2008one} for obtaining penalised estimates of the parameters in Generalised Linear Models (GLM) via the one-step sparse estimator by \cite{fan2001variable} for variable selection adopting nonconvace penalties; by \cite{McIlhagga2016} for penalized GLM based on Fisher scoring algorithms; by \cite{ZhuHuangReyes2010} for adaptive spatial LASSO in lattice data; and by \cite{ReyesEtAl2012} for penalised likelihood problems in linear spatiotemporal contexts. This study extends the aforementioned studies by obtaining penalised estimates of the fixed-effects coefficients of a linear mixed model for functional data, with the spatiotemporal dynamics modelled by a geostatistical random component.

We now consider the penalty function $f(\beta)$ in Equation (\ref{eq:adaptiveLASSO_0}). Motivated by the oracle properties of the adaptive LASSO estimates \citep{zou2006adaptive,BondellEtAl2010}, we use an adaptive penalty for the likelihood of the functional HDGM. Because the observed data are supposed to be correlated in space and time (due to the natural ordering of the spatiotemporal data), it is important that the algorithm be selection-consistent \citep{Zhang2010} even in the case of correlated observations. However, this may not often be the case for classic LASSO approaches \citep[see][among others, for conditions of selection consistency]{zhao2006model}. Thus, we suggest an adaptive LASSO penalty that has the desired property of selection consistency, as shown by \cite{zou2006adaptive,zou2008one,HuangHorowitzMa2008}.


We propose the following estimator with an adaptive LASSO penalty for the fixed-effects coefficients of f-HDGM:
\begin{equation}\label{eq:adaptiveLASSO}
 \beta^{(PMLE)}(\lambda) = \arg \min_{\beta} -\frac{1}{2} (\beta - \beta_0)' H_0 (\beta - \beta_0) + N \lambda \sum_{i=1}^{p} w_i |\beta_i| \quad ,
\end{equation}
where $\lambda$ is the regularisation parameter, and the penalty weights $w_i$ are chosen as the inverse initial ML estimates, that is, $w_i = \frac{1}{\|\beta_{0,i}\|^{\gamma}}$, with $\gamma = 1$ for all $i$. To increase or diminish the influence of these initial estimates, $\gamma \geq 0$ could also be chosen  differently. Generally, to obtain the oracle properties, the penalty parameter $\lambda$ should be of order $\sqrt{n}$ \citep[see][]{zou2008one}. In the next paragraph, we will go into further details on the selection of $\lambda$.

The algorithm used to solve Equation (\ref{eq:adaptiveLASSO}) is based on the BFGS quasi-Newton method over the non-zero coefficients, that is, the so called active set, with the initial values being $\beta_0$. The algorithm requires limited computation effort, as the time consuming computation of the Hessian matrix $H$ is done only once.

This penalised procedure shrinks irrelevant coefficients to zero. Because this applies to all basis functions separately -- we do not impose that all coefficients associated with one regressor must be shrunk to zero simultaneously as for a block LASSO approach --  it is possible to select the relevant sections of the functional coefficients and exclude the irrelevant knots. However, the basis functions may overlap to some extent. This implies that the height of the functional coefficient at a given point in the functional domain (i.e., the sum of the weighted basis functions at a given point) is determined by several coefficients. If only some of such coefficients are zero, the functional coefficient is not zero. Hence, typically, smooth transitions shrunken to zero can be observed in the functional domain, depending on the number and location of the knots (i.e., the fewer knots there are, the smoother the estimated function is). This further encourages the use of an adaptive LASSO penalty, which leads to asymptotically unbiased estimates \citep[see][]{zou2006adaptive}.

The penalty parameter $\lambda$ is determined by minimising the prediction errors obtained from a random K-fold cross-validation (CV) study. For this reason, let $\mathcal{D} = \{y_{s,t}(h), s \in S, t=1,...,T\}$ be the set of all available functional observations, and let $\mathcal{D}_1,...,\mathcal{D}_K$ be a random partition of $\mathcal{D}$, which is made by randomly assigning $N/k$ observation to each group $\mathcal{D}_i$ with $i=1,\ldots,K$.

For each subset $\mathcal{D}_i$, the penalised estimation is performed for a certain predefined sequence of $\lambda$, including when $\lambda = 0$. In particular, the parameters are estimated according to Equation \eqref{eq:adaptiveLASSO} using data in $\mathcal{D}-\mathcal{D}_i$. Then, the data in $\mathcal{D}_i$, are used to evlauate the out-of-sample prediction performance given by the root-mean-square error (RMSE$_i$) and the mean absolute error (MAE$_i$).

Eventually, the overall performance measures, say $RMSE(\lambda)$ and $MAE(\lambda)$, are obtained by averaging $RMSE_i$ and $MAE_i$ for $i=1,2,\ldots,K$. The optimal $\lambda^*$ may be chosen by minimising one of the following four CV criteria: $\arg \min_\lambda RMSE(\lambda)$, $\arg \min_\lambda MAE(\lambda)$ and the two corresponding \textit{one-standard-error rules}  \citep[see][]{HastieESL2017,HastieSLS2015}.
%
In the final step, the $\beta$ vector is re-estimated with all observations $\mathcal{D}$ at the optimal penalty parameter $\lambda^*$.

The procedure is synthesised using the pseudo-code in Algorithm \ref{fig:algorithm}.

\begin{algorithm}
	\caption{Adaptive LASSO estimation for functional hidden dynamic geostatistical models}\label{fig:algorithm}
	\begin{algorithmic}[1]
	\State Estimate the f-HDGM model parameters $(\beta_0)$ and the corresponding Hessian matrix $H_0$
	\For {Each partition $\mathcal{D}_i$ with $i=1,\ldots,K$}
        \For {$\lambda = 0, \ldots, \lambda_{max}$ on an exponentially decaying grid}
            \State Estimate $\beta^{(PMLE)}(\lambda)$ in Equation (\ref{eq:adaptiveLASSO}) using $\mathcal{D}-\mathcal{D}_i$ and initialise it with $\beta_0$
            \State Compute the fitted values (kriging) using $\beta^{(PMLE)}(\lambda)$
            \State Compute the $RMSE_{i}(\lambda)$ and $MAE_{i}(\lambda)$ using the observations in $\mathcal{D}_i$
		\EndFor
    \EndFor
    \State Compute $RMSE(\lambda)$ and $MAE(\lambda)$ and select $\lambda^*$ according to one of the four CV criteria
    \State Estimate the full model with $\lambda^*$ on $\mathcal{D}$ to obtain the final parameters $\hat{\beta}_{\lambda^*}$
	\end{algorithmic}
\end{algorithm}


\section{Monte Carlo simulation study}\label{sec:MC}
To evaluate the performance of the model selection algorithm, we performed a Monte Carlo simulation study based on three settings, labelled as \textit{Setting I}, \textit{Setting II}, and \textit{Setting III}. The three schemes are summarised in Table \ref{Tab:sim_schemes}. The settings represent the following situations of interest in geostatistical models. First, we regarded a multiple regression model as a benchmark approach (i.e., all temporal and spatial dependence parameters are chosen such that the resulting process is independent in space and time). Second, we considered the case of a dependent variable that is correlated across space and time but with uncorrelated regressors. Third, cross-correlation among the regressors was considered in \textit{Setting III}, which became the most challenging for model selection, but also the most realistic one. In particular, we consider cross-correlation ranging from moderate (0.5) to strong (0.9). The spatial dependence is exponentially decreasing with $\theta = 50$ km, so the correlation is below 0.37 after a distance of 50 km. The temporal autoregressive coefficients in the $G$ matrix are all equal to 0.85, resulting in a pronounced temporal persistence.

To represent a realistic spatial setting, we took the coordinates from the data that is used in the following empirical sections. More precisely, the coordinates of the spatial locations refer to the atmospheric monitoring sites that belong to the network of \textit{ARPA Lombardia}, the regional agency in charge of air quality monitoring in Lombardy. Its network consists of 84 ground stations distributed over its regional territory \citep[see the paper by][for an overview on ARPA Lombardia's monitoring system, the specifications of the measurement stations and available data provided by the region agency]{Maranzano2022}. Regarding the temporal resolution, we considered that the data were observed over 365 days, with each day represents the functional domain of 24 hours. For each of these settings, 500 Monte Carlo replications were simulated with the same geographical set-up. Thus, in total, $365\times24 = 8760$ hourly data were simulated for 15 locations 500 times.

\begin{table}
\caption{Specification of the simulation settings.}\label{Tab:sim_schemes}
    \begin{scriptsize}
        \begin{tabular}{l p{0.22\textwidth} p{0.22\textwidth} p{0.26\textwidth}}
        	\hline
        	\textbf{} & \textbf{Setting I} & \textbf{Setting II} & \textbf{Setting III} \\
        	\hline
        	Description & Uncorrelated response and regressors & Spatiotemporal correlation and uncorrelated regressors & Spatiotemporal correlation and correlated regressors \\
        	Spatial locations $s$ & 15 & 15 & 15 \\
        	Days $t$      & $365$ & $365$ & $365$ \\
        	Functional domain     & $[0,24]$ & $[0,24]$ & $[0,24]$ \\
        	Total observations & $15 \cdot 365\cdot24$ & $15 \cdot 365\cdot24$ & $15 \cdot 365\cdot24$ \\
        	Covariates & 
        	$\textbf{X} \sim N_3(\textbf{0},\Sigma_X)$ & 
        	$\textbf{X} \sim N_3(\textbf{0},\Sigma_X)$ & 
        	$\textbf{X} \sim N_3(\textbf{0},\Sigma_X)$ \\
        	Var-cov matrix & 
        	$\Sigma_X = \left[ \begin{array}{ccc} 1 & 0 & 0 \\ 0 & 1 & 0 \\ 0 & 0 & 1 \end{array} \right]$ & 
        	$\Sigma_X = \left[ \begin{array}{ccc} 1 & 0 & 0 \\ 0 & 1 & 0 \\ 0 & 0 & 1 \end{array} \right]$ & 
        	$\Sigma_X = \left[ \begin{array}{ccc} 1 & 0.9 & 0.7 \\ 0.9 & 1 & 0.5 \\ 0.7 & 0.5 & 1 \end{array} \right]$ \\
        	Spline type & B-spline & B-spline & B-spline \\
        	Interior knots & 5 & 5 & 5 \\
        	Spline order & 3 (cubic) & 3 (cubic) & 3 (cubic) \\ 
        	Number of bases & 7 & 7 & 7 \\ 
        	$\beta$ coefficients & [1 1 1 1 0 0 0] & [1 1 1 1 0 0 0] & [1 1 1 1 0 0 0] \\
        	$\theta$ & 0 km & 50 km & 50 km \\ 
        	\textbf{G} & diag(0,0,0) & diag(0.85,0.85,0.85) & diag(0.85,0.85,0.85) \\
        	$\Sigma_\eta$ & diag(1,1,1) & diag(1,1,1) & diag(1,1,1) \\
        	$\Sigma_\epsilon$ & diag(1,1,1) & diag(1,1,1) & diag(1,1,1) \\
        	\hline
        \end{tabular}
    \end{scriptsize}
\end{table}

\begin{figure}[htbp]
	\centering
	\includegraphics[width=1\linewidth]{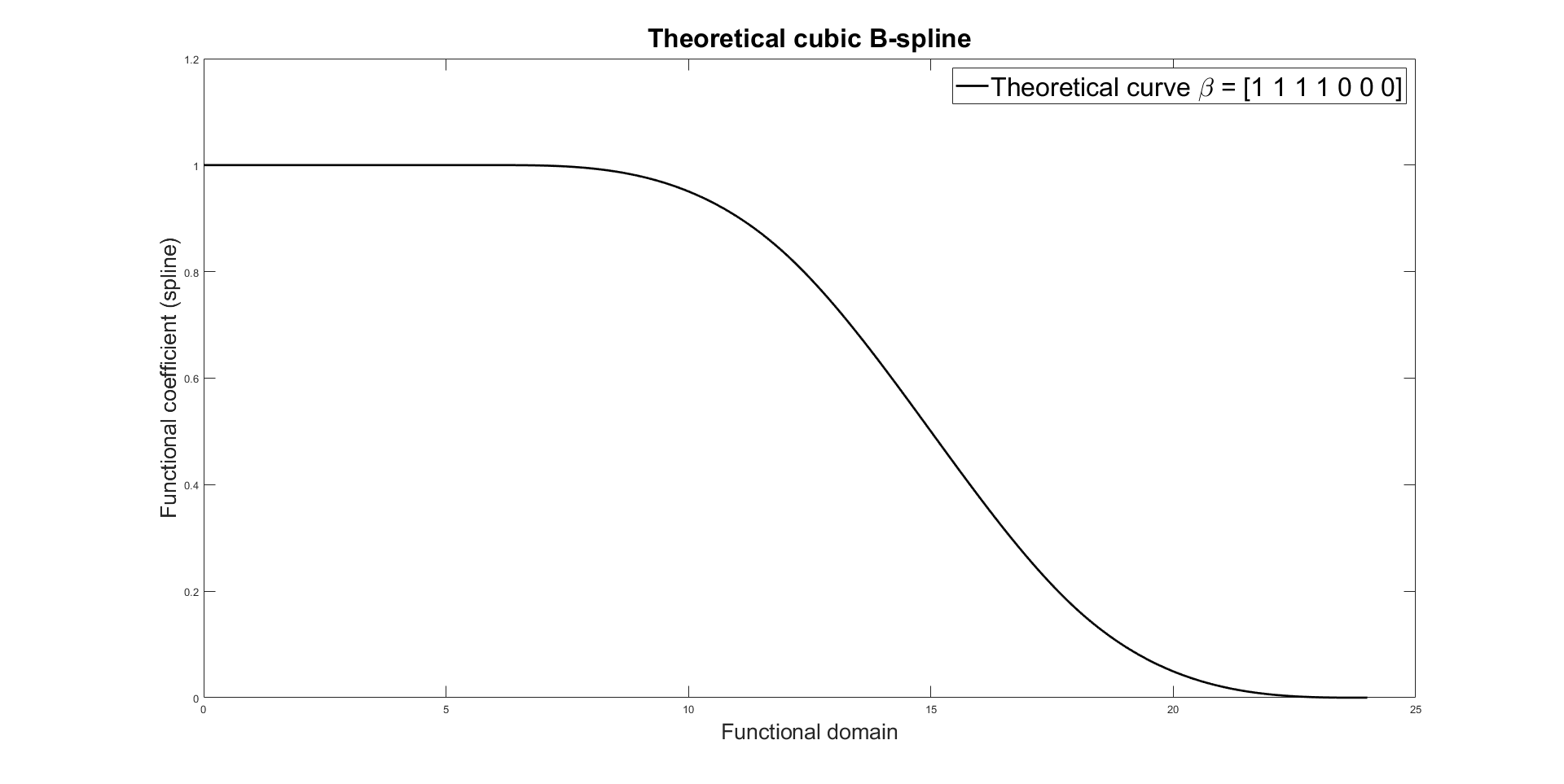}
	\caption{True functional coefficients with seven basis functions and coefficients equal $\beta = (1, 1, 1, 1, 0, 0, 0)'$ over a functional domain from 0 to 24.}
	\label{fig:TheorCubicSpline}
\end{figure}

For the functional interpolation, we considered a simple set-up of cubic B-spline basis functions with 7 knots. To analyse the performance of the algorithm in selecting relevant parts across the functional domain, we considered functional regression coefficients, that is 1 at the start of a day and going smoothly to 0, as shown in Figure \ref{fig:TheorCubicSpline}. Therefore, we set the coefficients of the first four bases equal to one, and to zero the remaining three.


The penalty term sequence was generated according to an exponentially decaying grid, starting from $\lambda_{min} = 10^{-5}$ up to $\lambda_{max} = 0.5$. As we will show in the simulation results, for a value of $\lambda$ greater than 0.50, all the penalised coefficients shrunk to 0. We added as the first value of the sequence the zero penalty, that is $\lambda = 0$, that corresponds to the unpenalised maximum likelihood solution. In total, we considered 101 different values. To identify the optimal value of $\lambda$, we performed a 10-fold random cross-validation across space and time.

For all the considered settings, we observed similar and reasonable results. For brevity, we will discuss only the detailed results of the spatiotemporal correlated observations with correlated covariates simulation setup, that is Setting III. The results of Setting I, Setting II and Setting III are reported in Appendix 3 to 5.

\begin{figure}[htbp]
	\centering
	\includegraphics[width=1\linewidth]{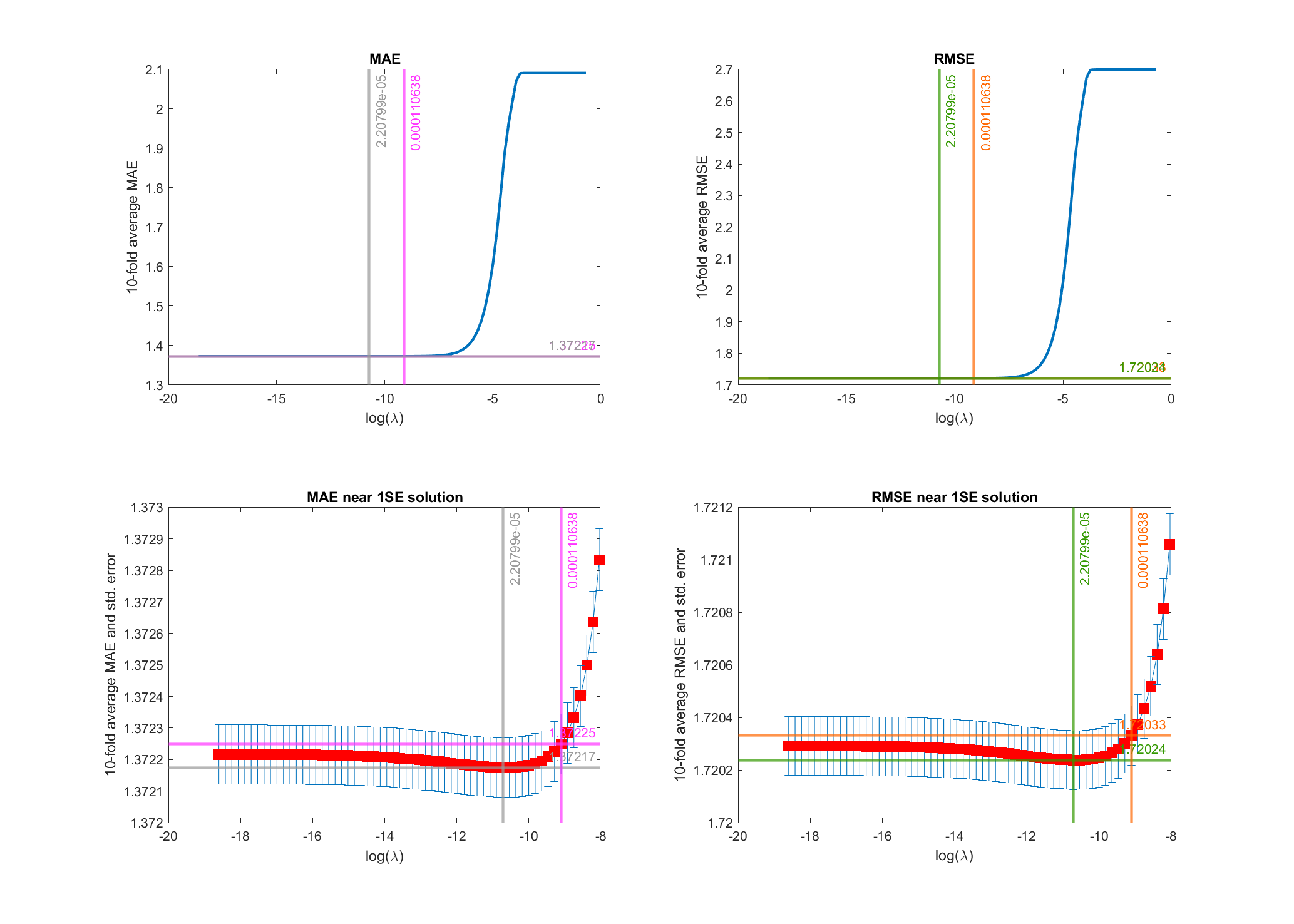}
	\caption{RMSE and MAE for different values of $\lambda$ in Setting III. Top panels: full $\lambda$ range. Bottom panels: near-optimum $\lambda$ range. Left panels: MAE. Right panels: RMSE. The vertical and horizontal lines correspond to the considered selection rules (grey: $\lambda^*_{MAE}$; pink: $\lambda^*_{1SE \ MAE}$; black: $\lambda^*_{RMSE}$; orange: $\lambda^*_{1SE \ RMSE}$).}
	\label{fig:STcorrCOVcorr_A}
\end{figure}

The major insights from the simulations are summarised as follows. First, Figure \ref{fig:STcorrCOVcorr_A} shows the average RMSE and MAE across the Monte Carlo replications and the optimal $\lambda$ values obtained after evaluating various criteria. Specifically, we considered the minimum error and one-standard-error from the minimum error. The upper panels show the MAEs (left) and the RMSEs (right) for all the values of $\lambda$ that were considered in this study, and the lower panels report the values that correspond to a reasonable range of $\lambda$ near the optimum solutions. Moreover, we depict the CV variability with the error bars, computed as the standard error of the sample average. Both the RMSE and MAE plots suggest that the optimal value of $\lambda$, that is, the minimiser of RMSE or MAE, is different from the MLE solution. Overall, both the RMSE and MAE showed smooth patterns. For the values of the penalty term $\lambda$ greater than 0.03 (i.e., $log(\lambda) > -3.50$), all the coefficients were shrunk to 0 and the RMSE and MAE stabilised around 2.7 and 2.1, respectively. Considering the one standard error rule, in both cases, the optimal values corresponded to $\lambda^*_{1 SE \ RMSE} = \lambda^*_{1 SE \ MAE} = 1.11\times10^{-4}$ (orange and pink vertical lines). However, the prediction performance did not significantly different from the MLE solution.

\begin{figure}[htbp]
	\centering
	\includegraphics[width=1\textwidth]{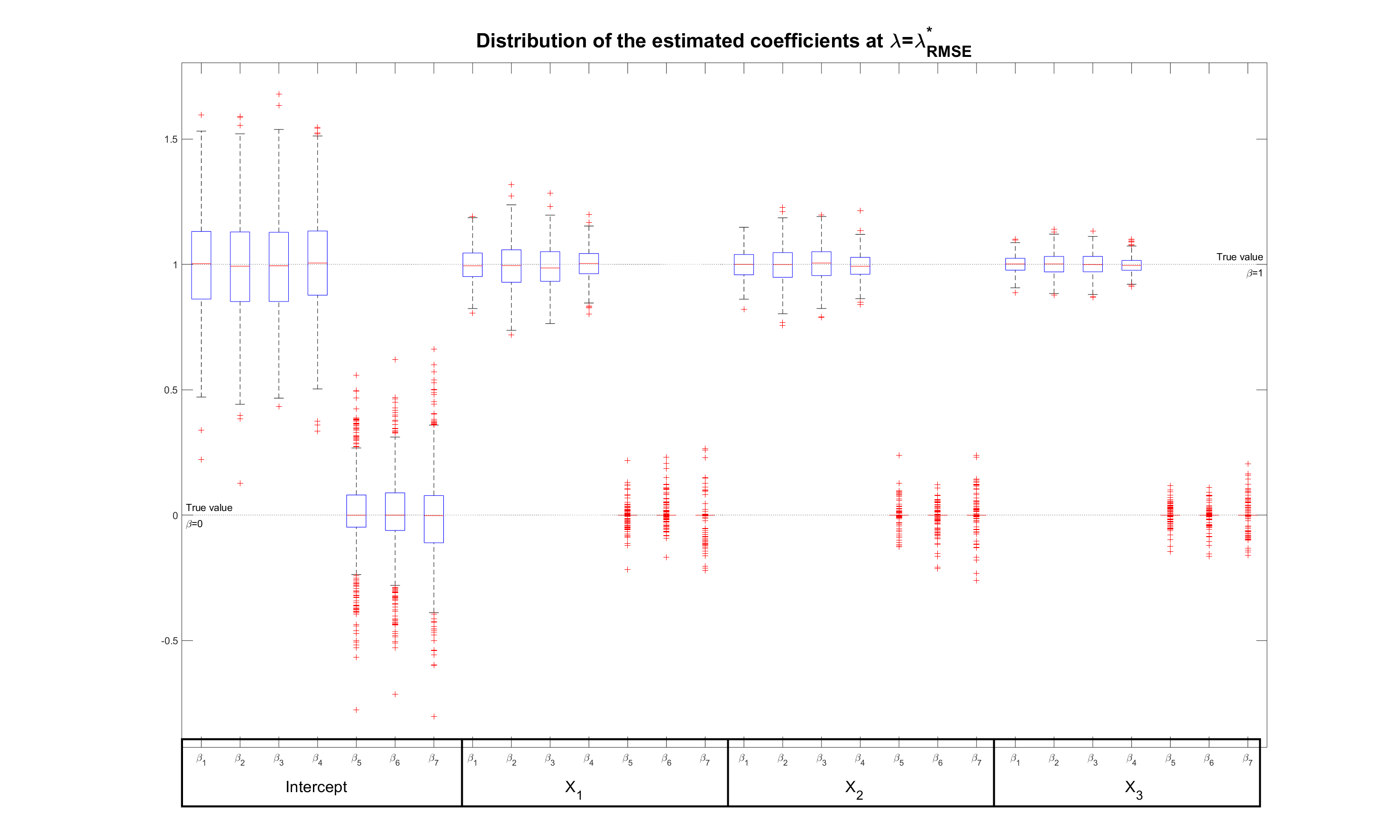}\\
	\includegraphics[width=1\linewidth]{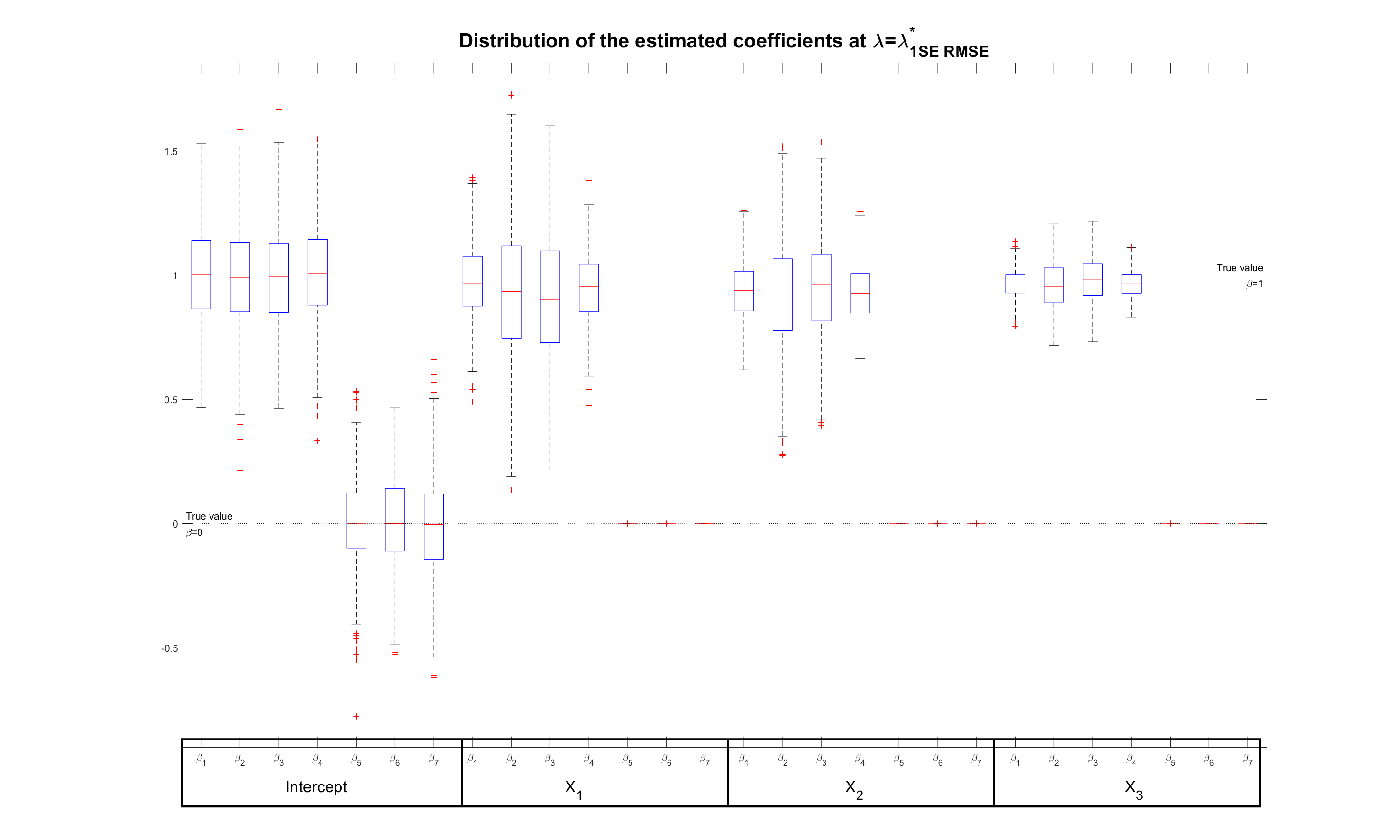}
	\caption{Box plot of the estimated coefficients across 500 simulations at $\lambda^* = \lambda_{min \ RMSE}$ (upper panel) and at $\lambda^* = \lambda_{1SE \ RMSE}$ (lower panel) for Setting III.}
	\label{fig:boxplot_beta_min}
\end{figure}

\begin{figure}[htbp]
	\centering
	\includegraphics[width=0.9\linewidth]{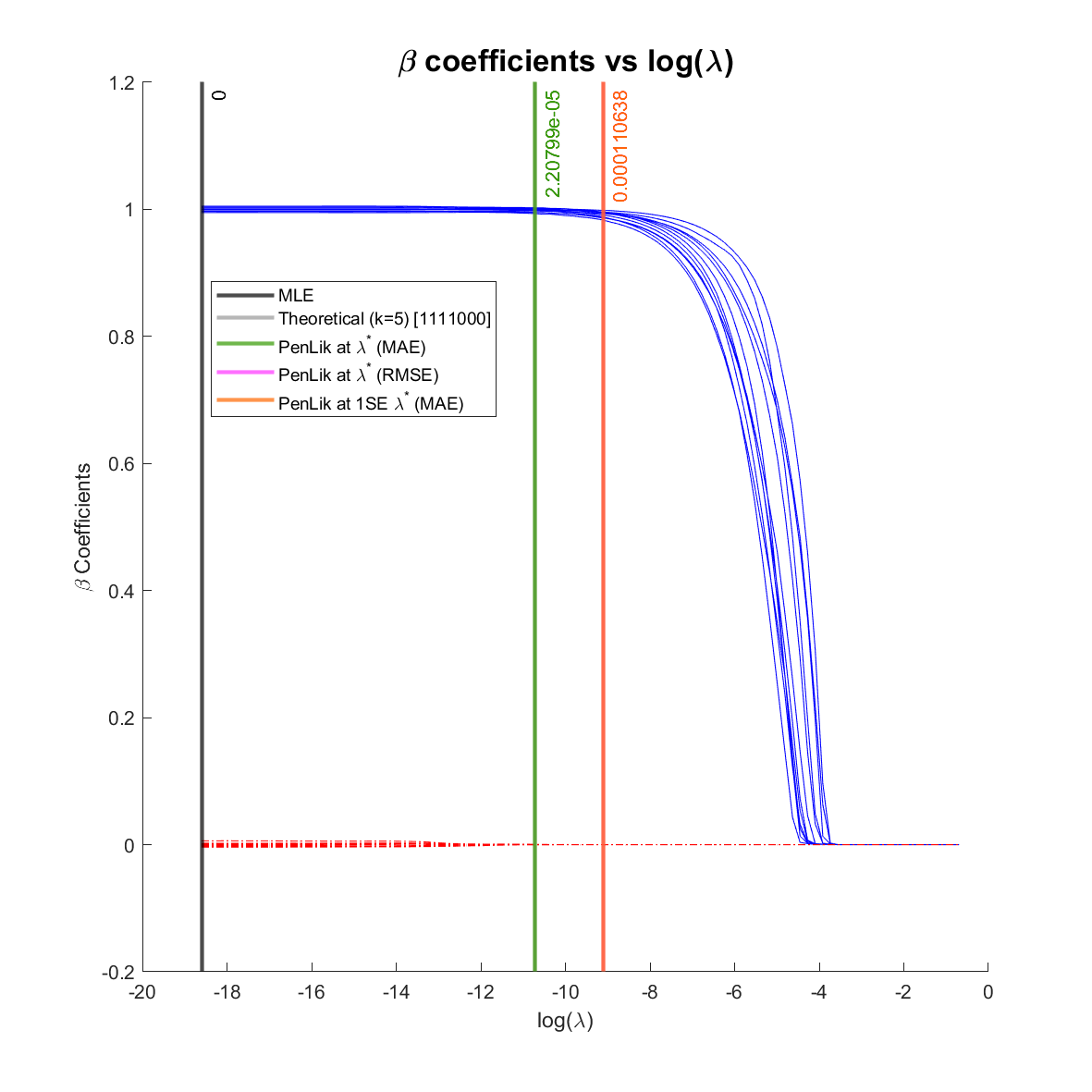}
	\caption{Average estimated coefficients for different values of $\lambda$ in Setting III. The positive coefficients are drawn in blue, and the zero coefficients are depicted by the red dashed lines.}
	\label{fig:STcorrCOVcorr_C}
\end{figure}

In Figure \ref{fig:boxplot_beta_min}, we plotted the empirical distribution (i.e., the box-plot) across the simulations of each fixed-effects coefficient for $\lambda = \lambda^*_{min \ RMSE}$ (upper panel) and $\lambda = \lambda^*_{1 SE \ RMSE}$ (lower panel). In both cases, the following are very noticeable: (1) the coefficients are estimated very close to their actual value, which indicates that the penalised estimators are approximately unbiased and (2) the variability for null coefficients is considerably smaller than for coefficients equal to 1. Similar considerations can be inferred from the further materials in Appendix 5. In particular Figure 11 reports the average functional pattern of each coefficient across the 24-hours, whereas Table 4, reports the average estimate and the sampling variability of each coefficients across the simulations.
Eventually, we can see from Figure \ref{fig:STcorrCOVcorr_C} that the average estimated coefficients are smoothly shrunk towards 0, where true zero coefficients are exactly 0 for $log(\lambda) > -3.5$, whereas the remaining positive coefficients are close to 1, which is their actual value.

\section{Application to air quality in Lombardy} \label{sec:AQLombardy}
\begin{figure}
	\centering
	\includegraphics[width=.8\linewidth,trim=0 0 0 50,clip]{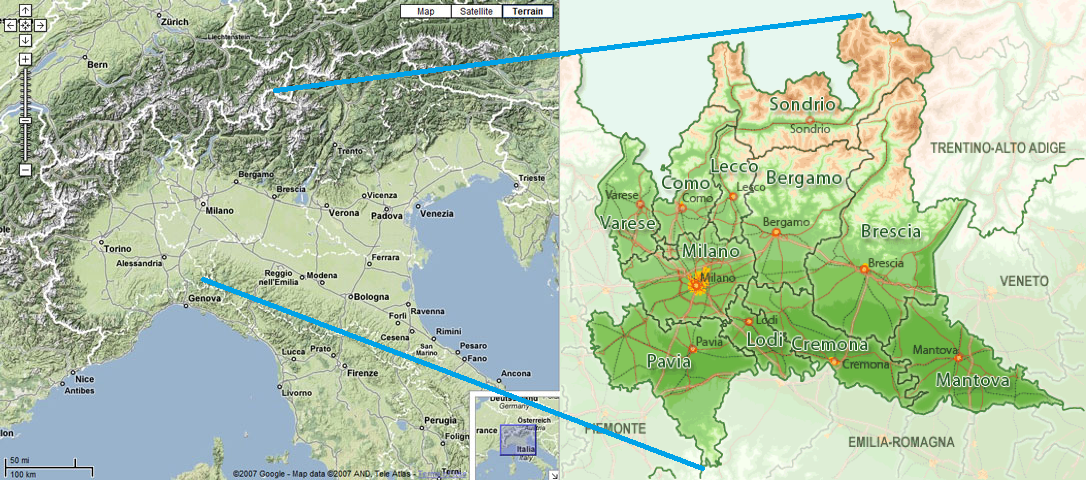}
	\includegraphics[width=.8\linewidth,trim=0 0 0 50,clip]{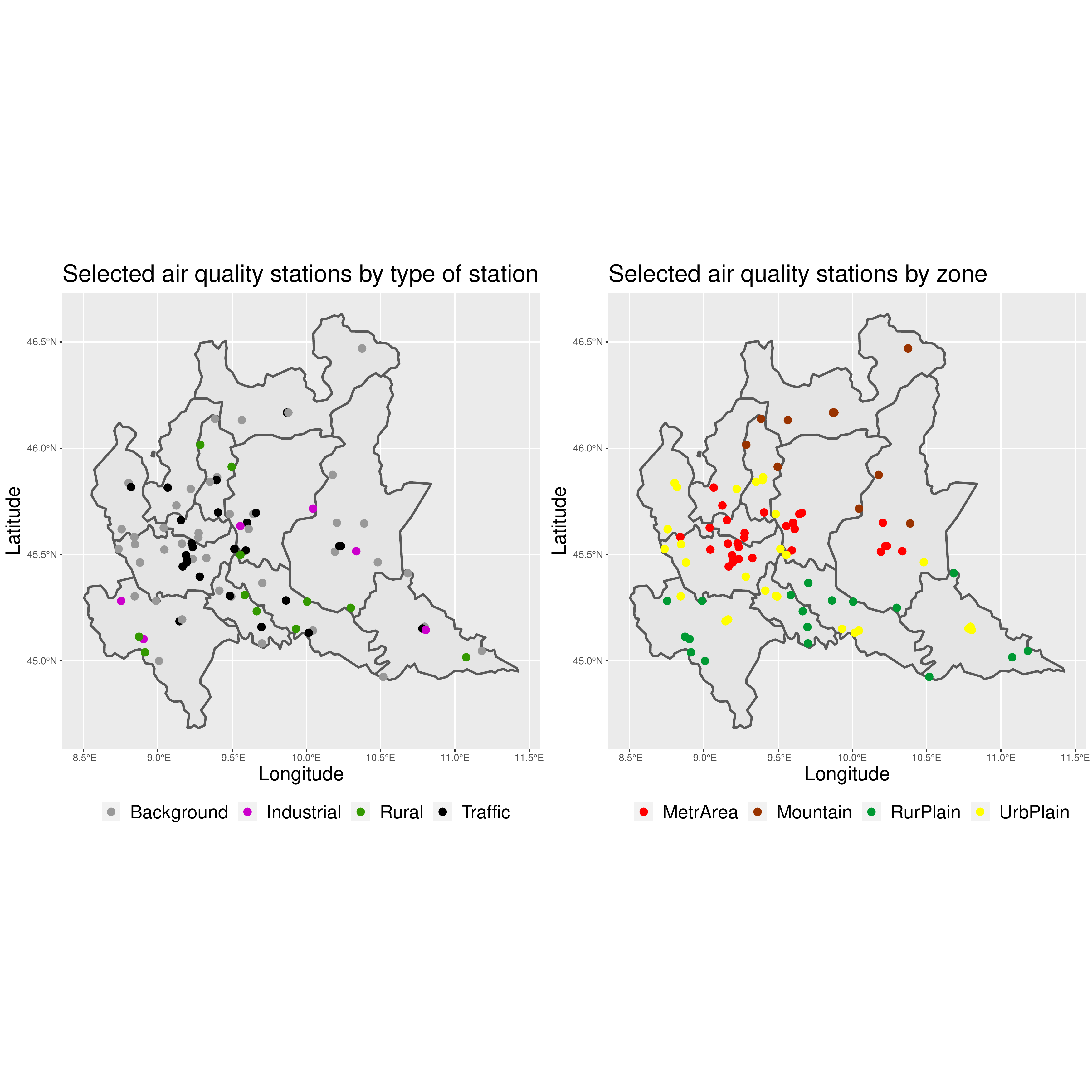}
	\caption{Physical map of Po Valley (upper left panel) and Lombardy (upper right panel) and the ARPA Lombardia air quality monitoring network by type of station (lower left panel) and type of area (lower right panel).}
	\label{fig:Lombardy}
\end{figure}

The proposed model selection algorithm is now applied to air quality data recorded during the COVID-19 pandemic in Lombardy (see Figure \ref{fig:Lombardy}). Airborne pollution in Lombardy has attracted considerable research interest for many years. With the COVID-19 emergence, many researchers have become increasingly interested in the short-term effects of lockdowns on air quality \citep{Cameletti2020,CollivignarelliEtAl2020,LovarelliEtAl2020,fasso2021spatiotemporal}. All the research results showed that the reduced mobility imposed by the government induced a positive effect on air quality, significantly reducing the concentrations of airborne pollutants directly related to traffic, such as nitrogen, particulate matters and benzene. However, particulate matter, did not appear to have been significantly reduced and persists at values similar to pre-pandemic levels \citep{Cameletti2020,GranellaEtAl2021,ARPA_COVID}.
This is seen to be because of the strong increases (up to +31\%) in the concentration of ammonia, the main source of emission of particulate matter, In the Lombardy countryside due to the non-interruption of agricultural activities therein \citep{LovarelliEtAl2021}

\begin{figure}
	\centering
	\includegraphics[width=1\linewidth]{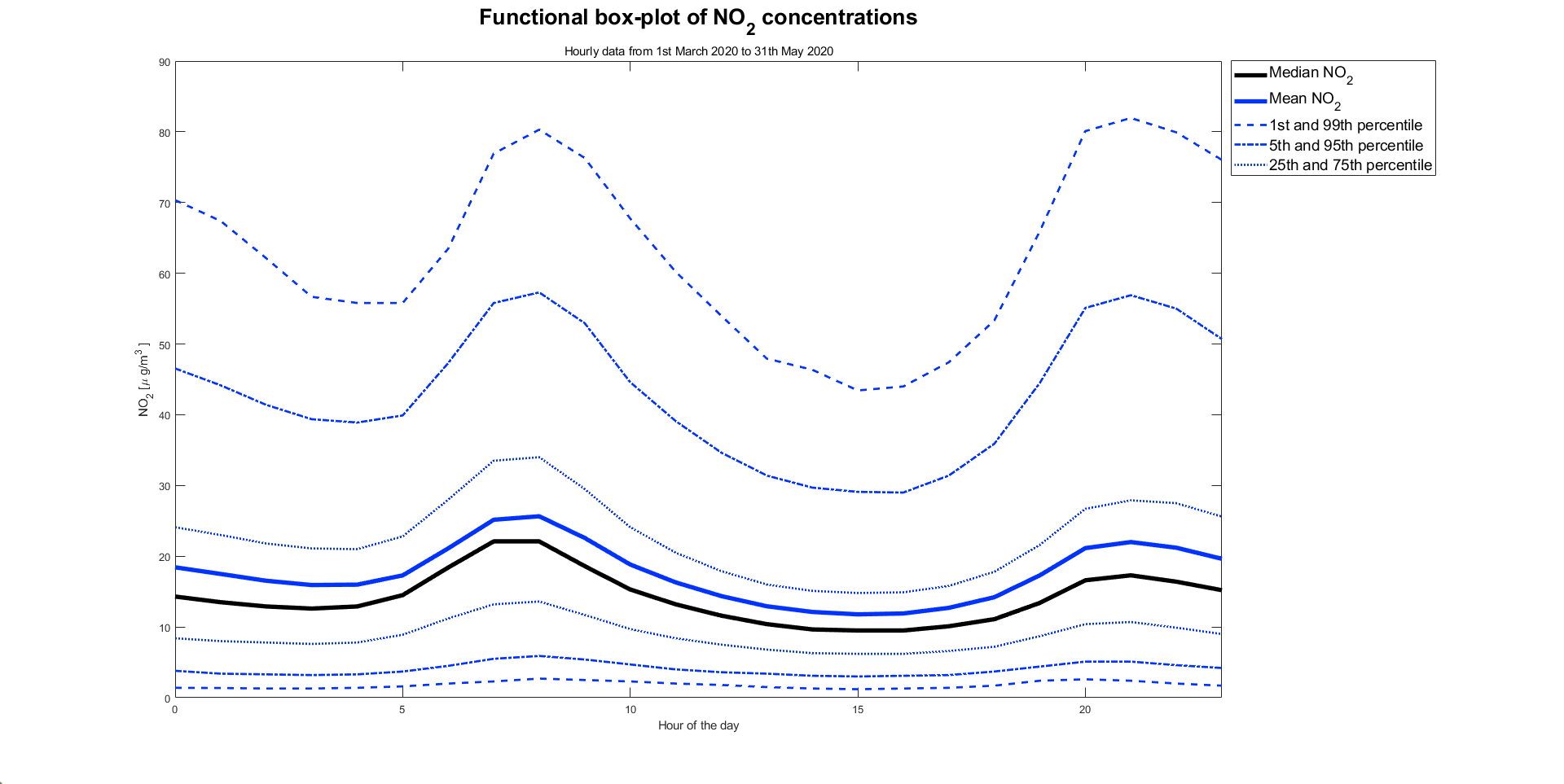}
	\caption{Intraday box-plot of NO$_2$ concentrations ($\mu g/m^3$) observed between the 1 March 2020 and 31 May 2020.}
	\label{fig:FunBox}
\end{figure}

We model hourly nitrogen dioxide (NO$_2$) concentrations obtained by the ARPA Lombardia monitoring network \citep{Maranzano2022} from $1$ March, 2020, to $31$ May, 2020, that is $T=92$ days. The monitoring network has $n=84$ ground stations, which are classified by type (background, industrial, rural and traffic stations) and zone (metropolitan, mountain, rural plain and urban plain areas). The station locations are shown in Figure \ref{fig:Lombardy}. Moreover, we consider the concentration throughout the day as a functional observation. For a first overview of the intraday variations, we show the regional daily distribution of NO$_2$ concentrations in the functional box plot in Figure \ref{fig:FunBox}. The 24-hour profile clearly shows the intra-day variability in the means and variances. In particular, there are strong differences between the NO$_2$ concentrations at night and day. They are in accordance with anthropogenic activities -- that is, very high concentrations are seen during rush hours, between 8 and 11 and between 17 and 23.

To explain the airborne pollutant concentrations, we considered a set of nine meteorological and land cover variables: temperature ($\mbox{}^{\circ} C$), precipitation (mm), relative humidity ($\%$), atmospheric pressure (Pa), eastward and northward component of the wind (m/s), geopotential height (m$^2$/s$^2$) and high and low vegetation covering  \citep[measured as one-half of the total green leaf area per unit horizontal ground surface area, cf.][]{ERA5Land}. Since the variables present different scales and ranges, we standardized both the response variable and the covariates with respect to their overall 24-hour mean and standard deviation. The total number of observations was $n \times T = 185472$ for each variable.

To account for the natural daily cycle, in Equation (\ref{eq:fHDGM}), we set $t$ as the day, whereas $h$ is the time across the day. Hence, periodic Fourier basis functions with $b$ bases and a support between 0 and 24 were used. Recall that by construction, Fourier splines require an odd number of bases, and their interpretation depends on the frequency of their calculation. In fact, except for the first basis, the following basis pairs were calculated at increasing seasonal frequencies. For example, if the number of bases was $b=5$, the pair formed by the fourth and fifth bases would have twice the frequency of the second and third pair. The use of Fourier bases ensures the continuity of the last hour of a day to the first hour of the consecutive day.
According to the number of basis functions, the total number of parameters to estimate is equal to $b \times 14$, In particular, $b \times 10$ parameters are associated with the covariates and the functional intercept; $b \times 3$ with spatiotemporal dynamics, and $b$ with residual component variances.

The algorithm was initialised by estimating the HDGM parameters, that is, the regression coefficients, the variance-covariance matrix and the spatiotemporal dynamics, using the unpenalised MLE. After estimating the full model, we applied the penalised likelihood model selection algorithm using an exponentially decaying grid of penalty coefficients $\lambda$ that ranged from $\lambda_{min}=10^{-4}$ to $\lambda_{max}=0.50$. We also included a value of $\lambda=0$ for the unpenalised estimates.

\subsection{Scalability of the algorithm}
We consider now the behaviour of our approach for increasing model complexity when applied to real world data. We are interested in computational costs and the algorithm behaviour.

To do this, we consider varying numbers of Fourier bases $b$ for each covariate and the related size of the variance-covariance matrix of the parameters. Also, we consider the impact of approximation methods for the fixed-effect coefficients (i.e., $\beta_0$) and the Hessian matrix (i.e., $H_0$) of the initial unpenalised f-HDGM. In particular, we consider spatial partitioning with groups varying from $k=1$ (no spatial partitioning) to $k=5$, and the approximated computation of the Hessian matrix, as described in Section \ref{sec:fHDGM}.

\begin{figure}
	\centering
	\includegraphics[width=1\textwidth]{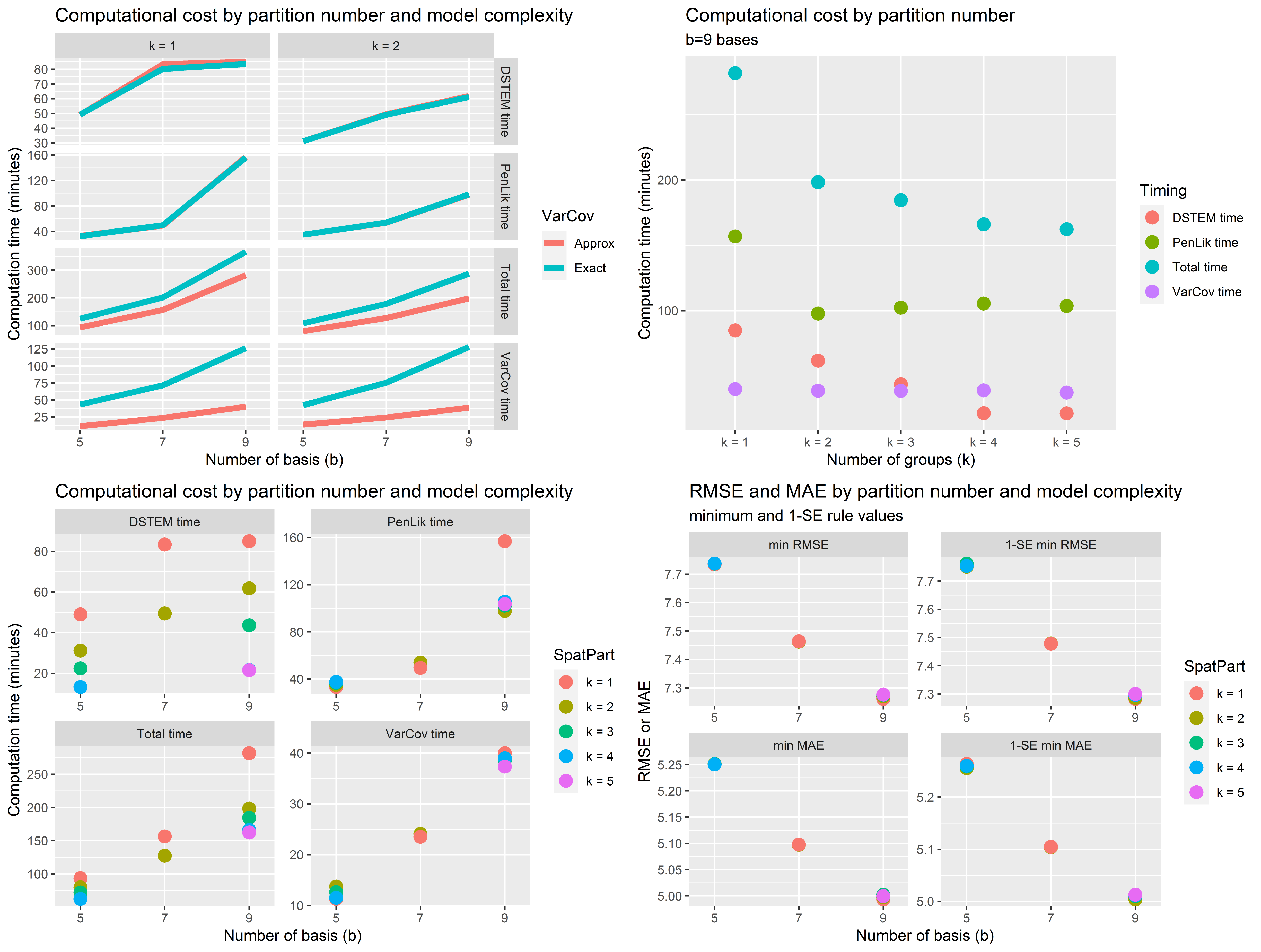}
	\caption{Computation time and cross-validation errors across the models. Computation time of each phase by model complexity with and without spatial partitioning (upper left panel); computation time of each phase by increasing level of spatial partitioning (upper right panel); computation time of each phase by increasing level of spatial partitioning and model complexity (lower left panel); RMSE and MAE by model complexity and increasing spatial partitioning (lower right panel).}
	\label{fig:comput_costs1}
\end{figure}

We want to examine the algorithm's ability to select only the relevant seasonal frequencies of the Fourier interpolation by shrinking irrelevant frequencies towards 0. Thus, we test the following occurrences as the complexity increases: (1) the increases in the number of fixed zero coefficients, and (2) the more frequent setting to 0 of the coefficients associated with high Fourier frequencies. These facts would be consistent with the observed values of the NO$_2$ concentrations (Figure \ref{fig:FunBox}), whose intra-day behaviour is fairly smooth and shows two peaks, which mean that from a modelling perspective, a small number of seasonal frequencies (low complexity) could be expected. A total of 17 models were evaluated. For each model, we considered the computation time for the three main phases of the algorithm -- that is, the initial model estimation with the EM algorithm, the computation of the variance-covariance matrix of the parameters, and the penalised likelihood algorithm. For model complexity we considered three scenarios: $b=5$, $b=7$ and $b=9$. In the first case, the total number of the model's parameters was $b \times 14 = $70; in the second case, is 98, and in the third case is $126$. The numbers of fixed-effect parameters were 50, 70, and 90, respectively.
%
The results are summarised in Figures \ref{fig:comput_costs1} and \ref{fig:comput_costs2}. The detailed results for all the models considered are given in Table 1 of Appendix 1.

\begin{figure}
	\centering
	\includegraphics[width=1\textwidth]{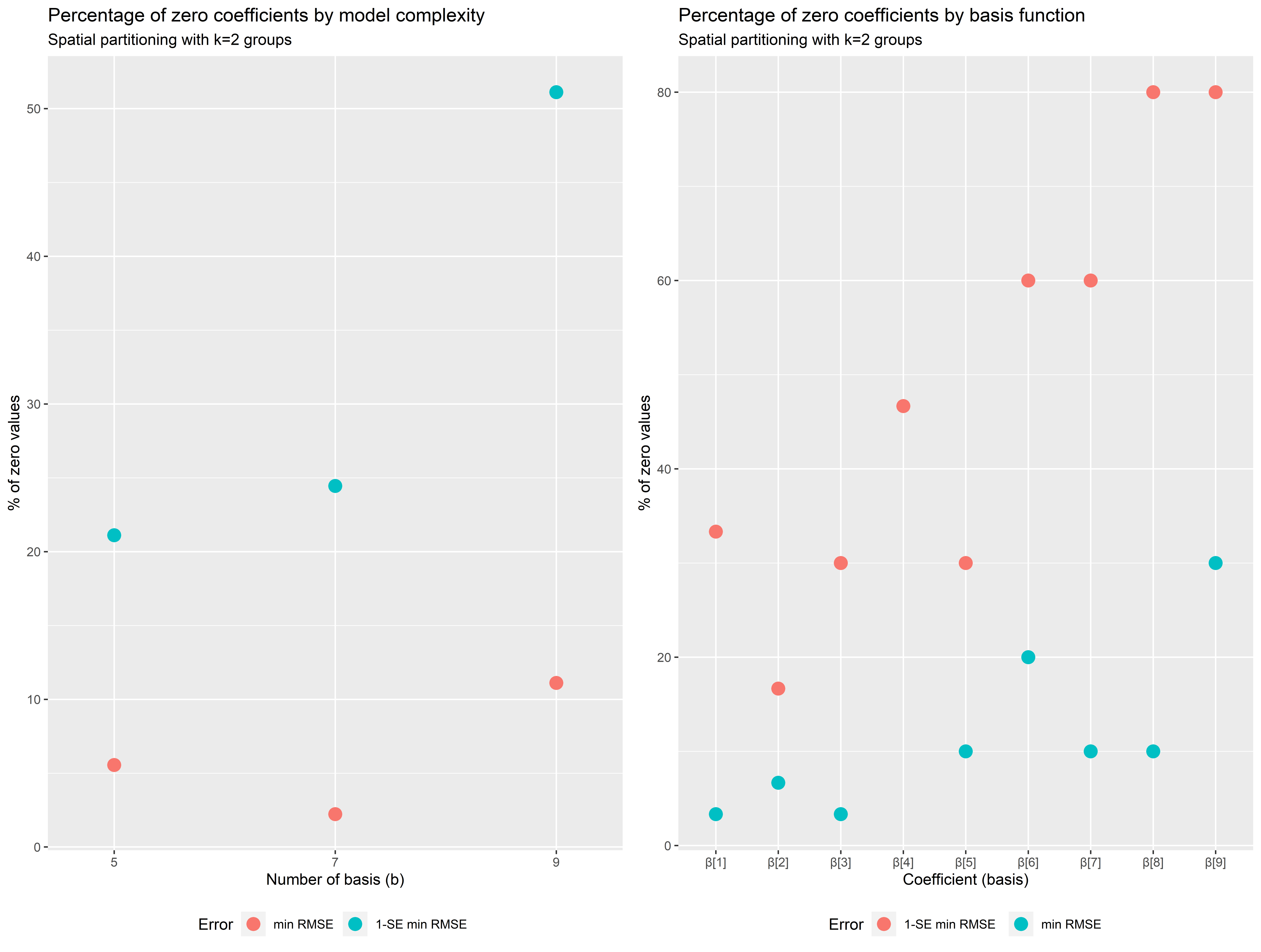}
	\caption{Percentage of zero coefficients across the models. Percentage of zero coefficients by model complexity when applying a spatial partitioning with $k=2$ groups (left panel); percentage of zero coefficients by basis function (coefficients) when applying a spatial partitioning with $k=2$ groups.}
	\label{fig:comput_costs2}
\end{figure}

Figure \ref{fig:comput_costs1} shows the computational costs of the penalised likelihood algorithm as a function of the model complexity, the spatial partitioning and the adoption of an approximation for the variance-covariance matrix. The main results are summarised as follows:

\begin{itemize}
    \item Variance-covariance matrix approximation (upper left panel of Figure \ref{fig:comput_costs1}): the approximated computation of the covariance matrix decreased its computation time of around 66\%, which in turns reduced the overall computation time by around 25\%. This holds independently from the model complexity. Of course, the penalisation algorithm is not affected.
    \item Spatial partitioning (upper right panel of Figure \ref{fig:comput_costs1}): the application of a spatial partitioning reduced the initial D-STEM computation time by 30\% to 50\% and the penalisation phase by up to 38\% . Moreover, it reduced the overall time of more than 30\%. The variance-covariance matrix computation was not affected. The time gain became negligible when the number of groups increased (i.e. $k \geq 4$).
    \item Model complexity (left panels of Figure \ref{fig:comput_costs1}): when $b$ was reduced from 9 to 7 basis functions, the computation time of all the phases significantly increased, independent of the approximation of the variance-covariance matrix. In particular, the penalised likelihood estimation decreased by up to 68\% and the overall computation by 45\%. When $b=5$ instead of $b=7$, the gain was still evident but less pronounced.
\end{itemize}

Concerning the cross-validated model error, both MAE and RMSE were affected only by the model complexity (lower right panel of Figure \ref{fig:comput_costs1}). Indeed, independent of the approximation of the covariance matrix or of the spatial partitioning, both MAE and RMSE decreased as the number of basis functions increased for all four criteria used to define the optimum $\lambda^*$.
In Figure \ref{fig:comput_costs2} we present the relationship between model complexity and the share of regressors removed with the adaptive LASSO. Previously, we stated that at an increasing number of fixed-effect coefficients, the number of irrelevant coefficients would increase and thus, would be set to 0 by the algorithm. Figure \ref{fig:comput_costs2} clearly shows that this statement is correct. Indeed, when the number of basis functions is large, the overall proportion of zero coefficients increased up to 25\% of the total. The graph on the right in Figure \ref{fig:comput_costs2} examines the excluded coefficients in detail and shows that the highest frequencies (corresponding to $\beta_6$ to $\beta_9$) were the most frequently removed by LASSO. This is consistent with what was observed with the NO$_2$ concentrations: since the response variable exhibited a very smooth intra-day pattern, the number of seasonal frequencies required to model the relationship with the covariates was significantly reduced. This result allows us to state that the adaptive LASSO algorithm proposed in this study can be a very useful tool for identifying the most relevant frequencies as it is precise in its selection and implementable even in contexts with large data sets. If time computing time is not an issue, using a higher number of frequencies (e.g., $b=9$ in our case) provides better forecasting performance (i.e., lower RMSE and MAE) while avoiding an excessive number of non-zero coefficients.

\subsection{Penalised estimates}
The numerical estimates showed very limited variability across the models and exhibited weak sensitivity to the approximations used in the estimation (i.e., spatial partitioning and Hessian matrix approximation). As shown in the preceding section, the impact of our adaptive LASSO procedure became more pronounced as the number of fixed-effects coefficients increases. Thus, we report and comment on the empirical results obtained considering the model with the lowest prediction error among those using $b=9$ basis functions. Specifically, we consider the case with an approximate variance-covariance matrix and without spatial partitioning.

\begin{figure}
	\centering
	\includegraphics[width=1\linewidth]{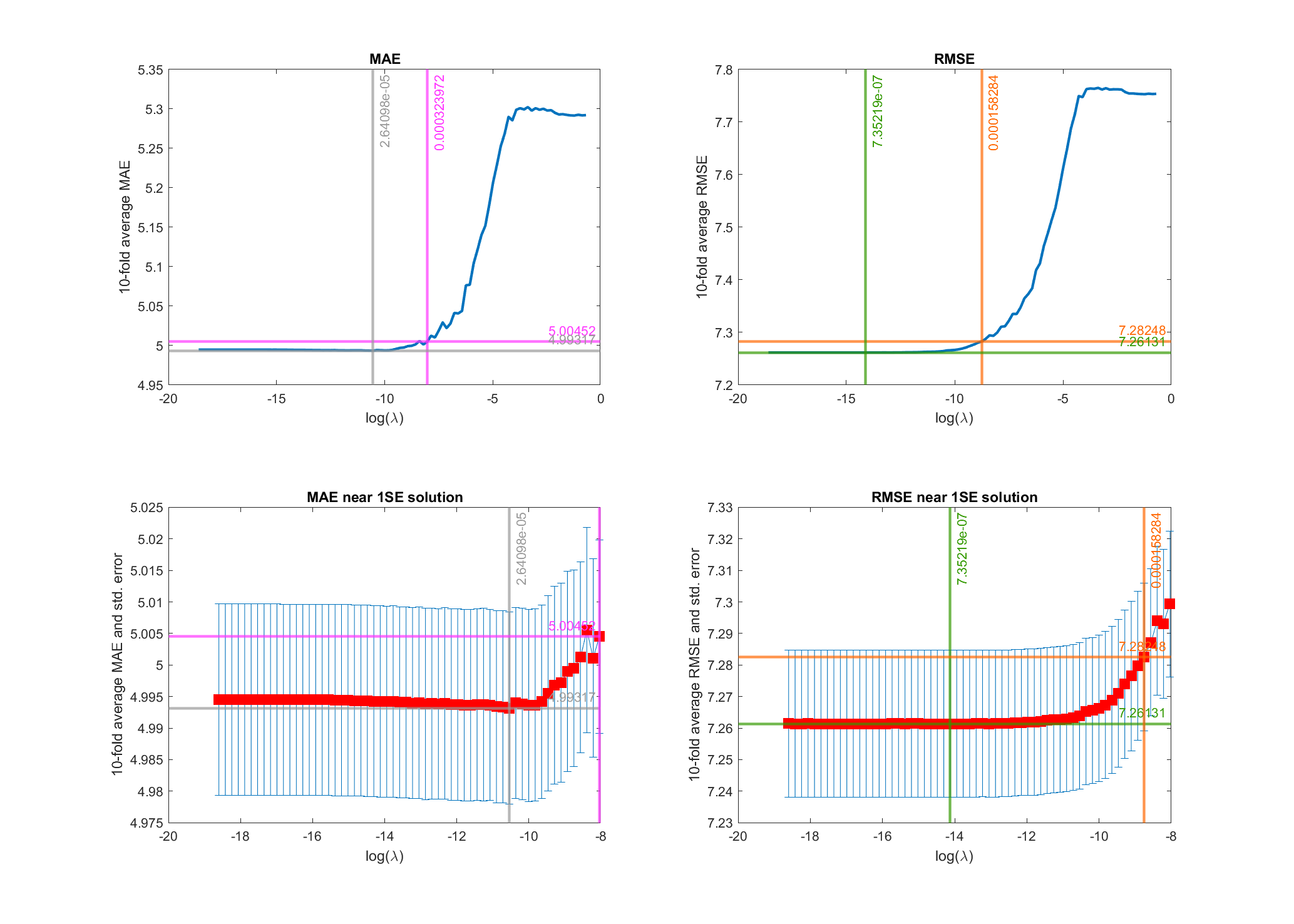}
	\caption{MAE against the logarithm of the penalty term $\lambda$ (left panels) and RMSE against the logarithm of the penalty term $\lambda$ (right panels). The horizontal lines represent the values of MAE and RMSE for key values of $log(\lambda)$, the optimal ($\lambda^*_{RMSE}$ and $\lambda^*_{MAE}$), 1-SE optimal values ($\lambda^*_{1SE \ RMSE}$ and $\lambda^*_{1SE \ MAE}$). The bottom panels are details near the optimum.}
	\label{fig:app_A}
\end{figure}

\begin{figure}
	\centering
	\includegraphics[width=1\linewidth]{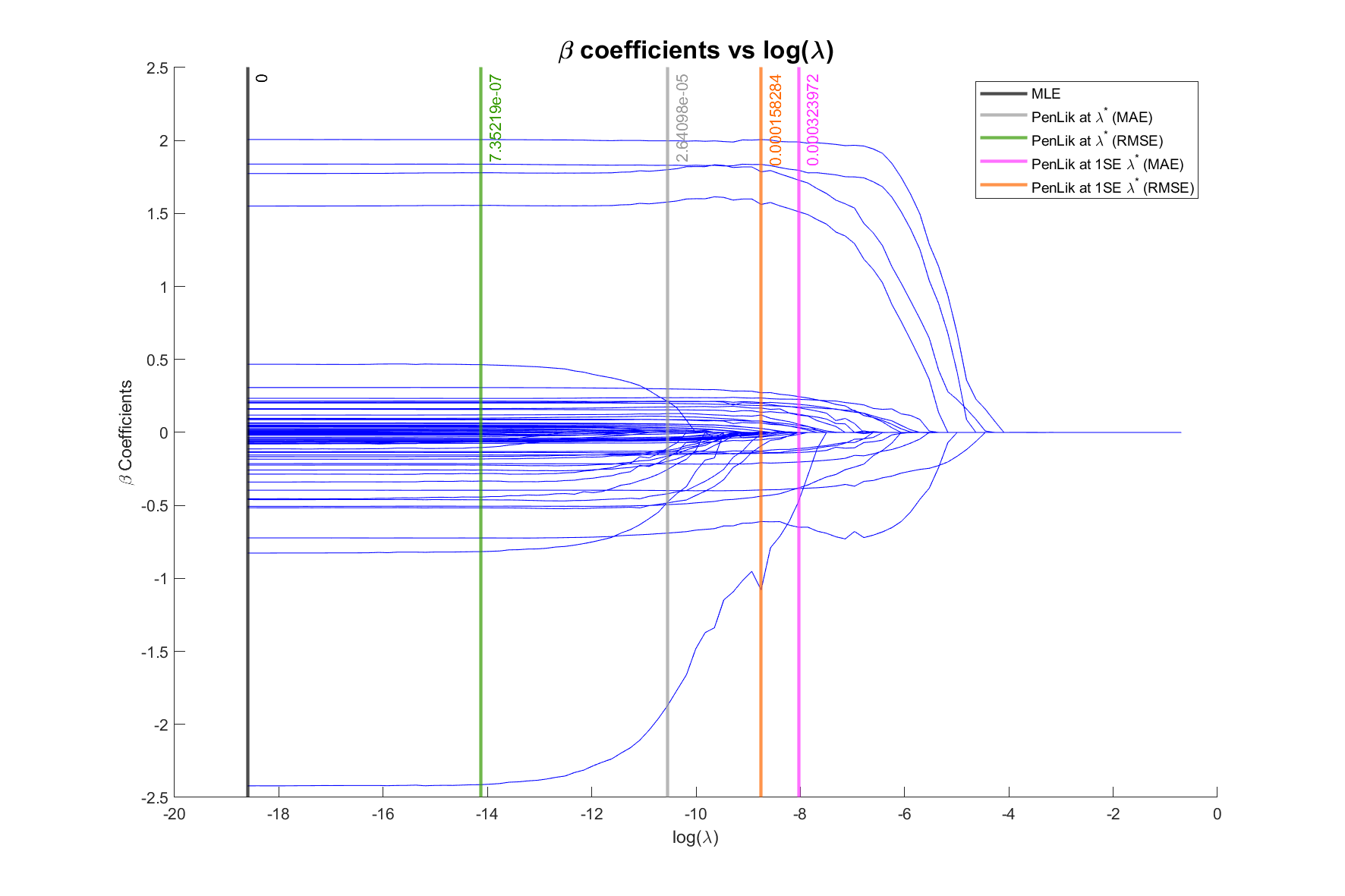}
	\caption{Functional coefficients against the logarithm of the penalty term $\lambda$. Vertical lines represent key values of $log(\lambda)$, i.e. unpenalized ($\lambda=0$), optimal ($\lambda^*_{RMSE}$ and $\lambda^*_{MAE}$), 1-SE rule ($\lambda^*_{1SE \ RMSE}$ and $\lambda^*_{1SE \ MAE}$).}
	\label{fig:app_C}
\end{figure}

In Figure \ref{fig:app_A}, we depict the behavior of the average cross-validation MAE (left panels) and RMSE (right panels) for increasing values of the penalty term $\lambda$. Both criteria highlight improvements in prediction errors when the adaptive LASSO was used to estimate the parameters. Both MAE and RMSE show a monotonic pattern apart from little sampling variability; and for large values of the penalty term (i.e., $log(\lambda) > -4$) all the covariates were dropped.
However, the one-standard-error-rule provided more parsimonious models with prediction errors not significantly different from the optimal ones. The estimated coefficients that correspond to the unrestricted model and the models associated with the four penalty terms considered are shown in Figure \ref{fig:app_C}.

\begin{figure}
	\centering
	\includegraphics[width=1\linewidth]{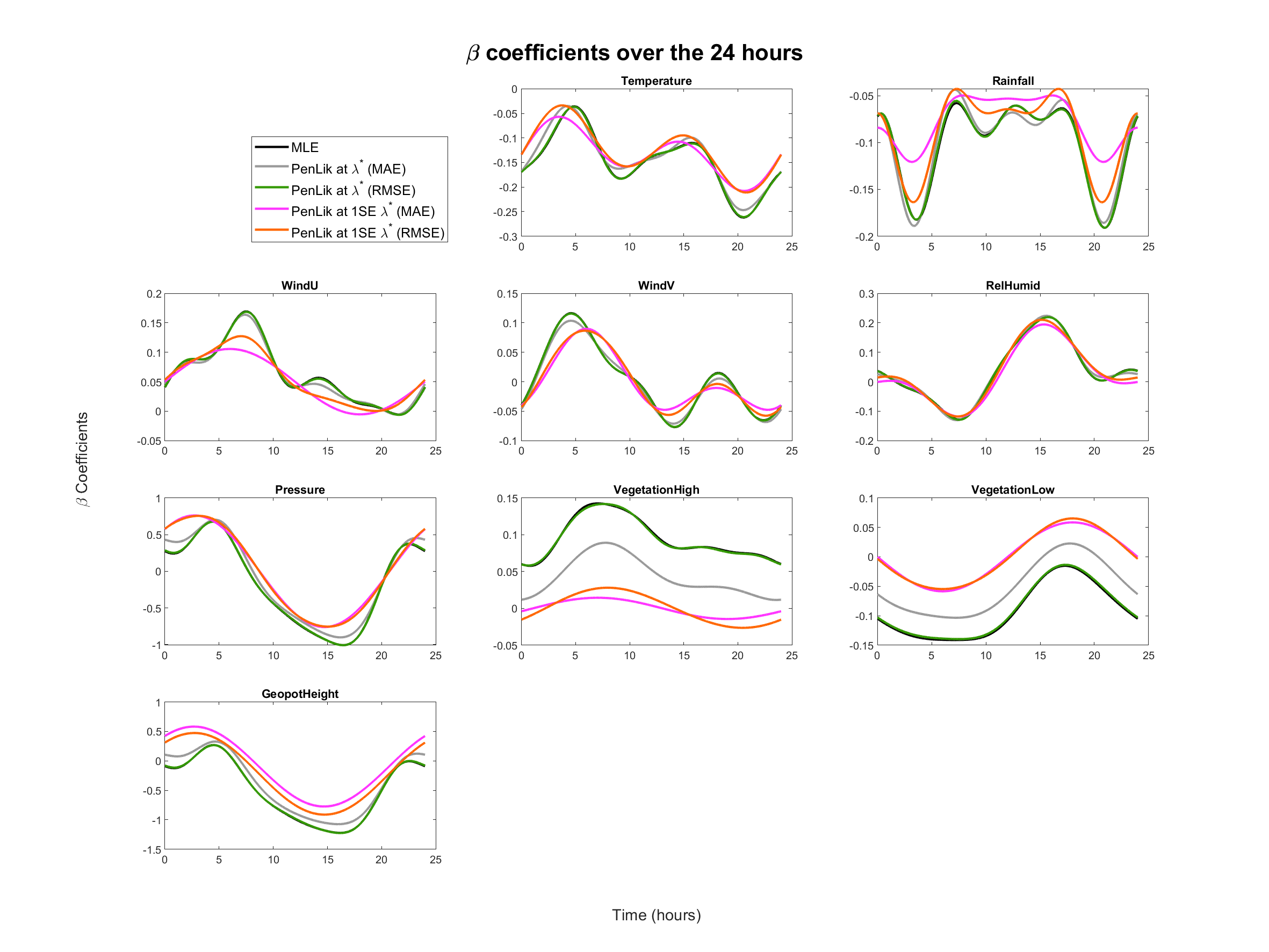}
	\caption{Estimated functional $\beta$ coefficients for differently selected optimal penalty parameters.}
	\label{fig:app_B}
\end{figure}

In Figure \ref{fig:app_B}, we show the 24-hour estimated functional coefficients for each variable. The black lines correspond to the unpenalised ML solution; the green lines to the optimal $\lambda$ w.r.t RMSE; the grey lines, to the optimal $\lambda$ w.r.t MAE; the orange lines, to the 1SE optimal $\lambda$ w.r.t RMSE; and the pink lines, to the 1SE optimal $\lambda$ w.r.t MAE. The estimated coefficients associated with temperature always exhibited negative values, particularly in the late afternoon and evening hours. The patterns obtained for different values of $\lambda$ did not show large discrepancies and tended to overlap throughout the day, leaving the overall dynamics unchanged during the day. However, the penalty seemed to mitigate the temperature effect at peak hours (10 a.m. and 8 p.m.).
Rainfall always showed a negative effect on the NO$_2$ concentrations, especially in the evening and just before dawn. In both moments, the effect reached the minimum peaks. Unlike the temperature, whose daily pattern varied slightly as the penalty increased, for higher values of $\lambda$, the precipitation diminished its effect and tended to flatten slightly towards 0. Considering the one-standard-error rule, between 7 a.m. and 5 p.m., the curve flattened to a constant negative value without being exactly 0. For the same $\lambda$ values, the two negative peak periods were greatly mitigated. These elements confirm the important role of temperature and rainfall in mitigating NO$_2$ concentrations, which is highlighted in literature \citep[e.g.,][]{fasso2021spatiotemporal}. Relative humidity presented some very interesting findings. First, its effect was null at around midnight, slightly negative at night before dawn and strongly positive in the daylight hours. Moreover, the penalisation appeared to produce no effect on the intra-day behaviour. This is consistent with the fact that whereas temperature and rainfall showed more complex patterns during the day, relative humidity already exhibited a simple pattern and did not need further smoothing.

Both the effects of atmospheric pressure and geopotential height (used as a proxy of elevation) depended on the moment of the day that was being considered. In both cases, the estimates showed a positive effect at the start and at the end of the day and a negative effect in the afternoon. However, in the case of elevation, the functional coefficient in the early and late hours was very close to zero, and in the central hours, it deviated significantly from 0 regardless of the penalty used. For both variables, penalisation did not play a significant role, the difference between penalised and non-penalised curves approached 0 and the infra-daily dynamics seemed to be stable.

Also, the U (eastward) and V (northward) components of wind showed a time-varying behaviour across the day. In both cases, the effect on the NO$_2$ concentrations was positively estimated during the early stage of the day, especially between 5 a.m. and 10 a.m., and it strongly weakened in the afternoon and at night, reaching values very close to 0 between 3 p.m. and 8 p.m. The morning positive estimate indicated that winds from the East (the Adriatic Sea) and the North (the Alps) reduced NO$_2$ concentrations, while those from the West and the South had a stagnating impact on the local NO$_2$ concentrations. However, the cleaning effect was limited to the early part of the day.
The shrinkage effect induced by the penalisation algorithm was more pronounced in the eastward component than in the northward component. In fact, we noticed that the effect of the eastward component was strongly smoothed in the morning hours, and the coefficient was cancelled during the afternoon. The northward component, although also smoothed, showed a significantly positive effect in the early hours of the day.

Finally, we saw that penalisation generated a remarkable influence on the two land cover variables, that is the high vegetation and low vegetation indices. Both variables were heavily squeezed towards 0 even for the contained values of $\lambda$ until they became almost 0 for the values associated with the one-standard-error rule. Similar results were presented in \cite{fasso2021spatiotemporal}, in which the effect of the same covariates on the NO$_2$, PM$_{10}$ and PM$_{2.5}$ concentrations in Lombardy was estimated to be close to 0, and thus not statistically significant, with the exception of the most urbanised areas.

\section{Conclusions and future developments}\label{sec:conclusion}
In this paper, we introduced an adaptive LASSO estimator for functional hidden dynamic geostatistical models. This new estimation approach based on penalised maximum-likelihood estimation can be used to efficiently select relevant covariates in functional geostatistical models.

Our proposal may be useful in environmental policy assessment. Special cases are agricultural policies considered by the mentioned Agrimonia project, air quality assessment \citep{fasso2021spatiotemporal}, traffic policy \citep{IJERPH2020},  sustainable development \citep{WangFangWang2016} and energy policies \citep{YuanShaohuaYannaTong2018}.

Moreover, the algorithm can be successfully applied to identify only relevant parts of the coefficients across the functional domain. From a computational perspective, we showed that the estimation can be efficiently implemented as a local quadratic approximation around the maximum of the expected likelihood function. To find this maximum, the EM algorithm implemented in the D-STEM software can be used  \citep[see][]{finazzi2020dstem2}. Then, we used a BFGS quasi-Newton iterative method to optimise the penalised function.

We analysed the performance of this estimation procedure through a Monte Carlo simulation study based on three settings with increasing level of complexity and representative of common applied contexts. To be precise, we considered settings where only parts of the functional coefficients had zero effects and where the regressors were cross-correlated and driven by spatiotemporal dynamics, as is often observed in geostatistical applications. Eventually, we applied the penalisation algorithm to an empirical example of air quality assessment. In particular, we modeled the $NO_2$ concentrations observed in Lombardy during the COVID-19 pandemic period in Spring 2020. In addition to showing the direct effects of the adaptive LASSO algorithm on the estimated coefficients, we provided a study of the scalability of the algorithm when applied to real world data. In particular, we showed that even with high model complexity, the computation time (both of the penalised likelihood and overall) can greatly benefit from approximations in model estimation, leaving performance essentially unaffected.

This paper focused on model selection in functional data contexts, performed using an adaptive LASSO penalisation algorithm. However, further extensions can be pursued. Indeed, the smoothness of the estimated functional coefficients can also be of high interest in such applications because too large a number of spline bases leads to over-fitting and non-smooth estimated effects. A further penalty term based on the integrated second derivatives could counter these effects. Thus, an elastic net structure that includes the smoothness penalty and an adaptive LASSO penalty is a very interesting topic for future research.
Eventually, using the results of \cite{SimonTibshirani2012}, the standardised group-LASSO estimator can be easily extended to spatiotemporal functional models by optimising a penalized likelihood function with quadratic approximation and assuming that the spline basis functions associated with each covariate are a group.


\section*{Data and codes}
All the results presented in this paper can be reproduced using Matlab software. Data and codes for both the simulation and application results are available at the following Google Drive link: \linebreak\href{https://drive.google.com/drive/folders/1de4Is3hw9davfo35evTOCUlbOTbPHS4i?usp=sharing}{https://drive.google.com/drive/folders/1de4Is3hw9davfo35evTOCUlbOTbPHS4i?usp=sharing}.


\section*{Declarations}
Funding: this research was funded by Fondazione Cariplo under the grant 2020-4066 "AgrImOnIA: the impact of agriculture on air quality and the COVID-19 pandemic" from the "Data Science for science and society" program.

\end{document}